\documentclass[twocolumn,superscriptaddress,aps,prx]{revtex4-1}
\usepackage{amsfonts}
\usepackage{amsmath}
\usepackage{amssymb}
\usepackage{graphicx}
\usepackage{color}
\usepackage{ulem}
\usepackage{dcolumn}
\usepackage{bm}
\usepackage{url}
\usepackage[colorlinks, urlcolor = cyan, citecolor = blue, linkcolor = magenta]{hyperref}
\usepackage{ulem}
\usepackage{lipsum}
\usepackage{diagbox}

\begin{document}

\title{Precision measurement for open systems by non-hermitian linear response}
\author{Peng Xu}
\email{physicalxupeng@whu.edu.cn} 
\affiliation{School of Physics, Zhengzhou University, Zhengzhou 450001, China}
\affiliation{Institute of Quantum Materials and Physics, Henan Academy of Sciences, Zhengzhou 450046, China}
\author{Gang Chen}
\email{chengang971@163.com}
\affiliation{Laboratory of Zhongyuan Light, School of Physics, Zhengzhou University, Zhengzhou 450001, China}

\begin{abstract}

    The lower bound of estimated precision for a coherent parameter unitarily encoded in closed systems has been obtained, and such a lower bound is inversely proportional to the fluctuation of the encoding operator. In this paper, we first derive some general results regarding the lower bound of estimated precision for a dissipative parameter, which is non-unitarily encoded in open systems, by combining the law of error propagation and the non-hermitian linear response theory. This lower bound is related to the correlation of the encoding dissipative operator and the evolution time. We next demonstrate the utility of our general results by considering three different kinds of non-unitary encoding processes, including particle loss, relaxation, and dephasing. We finally compare the lower bound with the quantum Fisher information obtained by tomography and find they are consistent in the regime where the non-hermitian linear response applies. This lower bound can guide us to find the optimal initial states and detecting operators to significantly simplify the measurement process.

\end{abstract}

\maketitle

\section{Introduction}

The linear response theory, which links measurement and correlation, is crucial for the interpretation of equilibrium properties of systems in areas such as condensed matter and cold atoms~\cite{Mahan1981Many, Coleman2015Introduction}. For example, the current detected by scanning tunneling microscopy reflects the spectroscopy of surfaces, while the measurement of angle-resolved photoemission spectroscopy reflects the spectroscopy of bulks. Besides, the Hall conductivity corresponding to the current-current correlation reflects the topology of materials~\cite{Coleman2015Introduction}. Moreover, the linear response theory also plays an important role in precision measurement~\cite{Hauke2016Measuring}. In the area of precision measurement, the estimation for a parameter $f$ is usually based on the law of error propagation $\Delta f = \langle \Delta \hat{O}(t) \rangle / |\partial \langle \hat{O}(t) \rangle / \partial f|$, where $\hat{O}(t)$ is an observable in the Heisenberg picture with mean value $\langle \hat{O}(t) \rangle$ and fluctuation $\langle \Delta \hat{O}(t) \rangle$. According to this law, we find the estimation for $f$ will be more precise if the fluctuation of $\hat{O}(t)$ is smaller and the response of $\hat{O}(t)$ corresponding to $f$ is larger. 

\begin{figure}
  \centering
  \includegraphics[width=8cm]{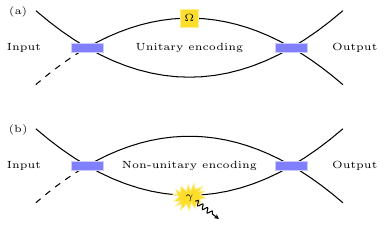}
  \caption{Schematics of two different kinds of interferometers for precision measurement. (a) Unitary interferometer. The parameter $\Omega$ is unitarily encoded in the system, $\hat{U} = e^{- i \hat{H}_1 t}$ and $\hat{H}_1 = \Omega \hat{A}, \hat{H}_1^\dagger = \hat{H}_1$. (b) Non-unitary interferometer. The parameter $\gamma$ is non-unitarily encoded in the system, $\hat{U} = e^{- i \hat{H}_\text{diss} t}$ and $\hat{H}_\text{diss} = - i \gamma \hat{A}^\dagger \hat{A} + \left( \hat{\xi}(t) \hat{A}^\dagger + \hat{A} \hat{\xi}^\dagger(t) \right), \hat{H}_\text{diss}^\dagger \neq \hat{H}_\text{diss}$.} 
  \label{fig:interferometer}
\end{figure}

In precision measurement, the Ramsey interferometer is usually used to estimate the parameter $\Omega$, as shown in Fig.~\ref{fig:interferometer}(a)~\cite{Yurke1986SU}. For this interferometer, the external field is unitarily encoded in the system as $\hat{H} = \hat{H}_0 + \hat{H}_1$, where $\hat{H}_0$ is the Hamiltonian of the system before encoding and $\hat{H}_1 = \Omega \hat{A}$ is the encoding part. Combining the law of error propagation and the linear response theory $\delta \langle \hat{O}(t) \rangle = - i \Omega \int_{0}^t dt' \langle [\hat{O}_\text{I}(t), \hat{A}_\text{I}(t')] \rangle$~\cite{Kubo1957Statistical}, we obtain the estimated precision for the parameter $\Omega$, $\Delta \Omega \geq 1 / 2 \langle \Delta \hat{A} \rangle t$, which is bounded by the fluctuation of the encoding operator and the evolution time, as demonstrated in Appendix.~\ref{sec:app1}. Hence, the entangled states and the critical states, since its large fluctuation for some encoding operators, have been widely used to enhance the metrology to the Heisenberg limit~\cite{Gross2010nonlinear, Riedel2010atom, Hamley2012Spin, Strobel2014Fisher, Muessel2014Scalable, Luo2017Deterministic, Xu2019Efficient, Bao2020Spin, Xin2023Long, Sewell2012Magnetic, Cox2016Deterministic, Bornet2023Scalable, Youssefi2023Squeezed, Robinson2024Direct}. For example, the spin squeezed state has been used to enhance the metrology for external magnetic fields~\cite{Colombo2022Time, Liu2022Nonlinear, Mao2023Quantum}, and the critical state has been used to enhance the metrology for external electric fields~\cite{Saffman2010Quantum, Facon2016Sensitive, Ding2022Enhanced, Gammelmark2011Phase, Macieszczak2016Dynamical, Fern2017Quantum, Raghunandan2018High, Garbe2020Critical, Chu2021Dynamic, Montenegro2021Global}. However, a natural and unresolved question is what determines the lower bound of the estimated precision for a dissipative parameter, when the system is coupled to white noise?

In this paper, we consider estimating the parameter $\gamma$, as shown in Fig.~\ref{fig:interferometer}(b), which is non-unitarily encoded in the system as $\hat{H} = \hat{H}_0 + \hat{H}_\text{diss}$. $\hat{H}_\text{diss} = - i \gamma \hat{A}^\dagger \hat{A} + \left( \hat{\xi}(t) \hat{A}^\dagger + \hat{A} \hat{\xi}^\dagger(t) \right)$ is the encoding dissipative part, where $\gamma$ is the strength of dissipation, $\hat{A}$ the loss operator, and $\hat{\xi}(t)$ the Langevin noise~\cite{Scully1997Quantum, Pan2020Non, Deng2021Stability}. We first analyze the scaling of estimated precision for the parameter $\gamma$ with quantum resources such as particle number $N$ and evolution time $t$, and demonstrate the estimated precision is bounded by the correlation of the encoding dissipative operator. Then, we explicitly calculate the estimated precision for three different kinds of non-unitarily encoding operators, including dephasing, relaxation, and particle loss. These numerical results further confirm the analytical results for the lower bound. Additionally, we also compare our results with the quantum Fisher information obtained by tomography, and we find they are consistent in the regime where the non-hermitian linear response applies. The lower bound we obtained based on the non-hermitian linear response theory may prove to be pivotal in finding the optimal initial states and detecting operators for precision measurements performed in open systems. 

\section{General formalism}

When a system is coupled to white noise, it can be described by an effective non-hermitian Hamiltonian
\begin{equation}
    \hat{H} = \hat{H}_0 + \hat{H}_\text{diss},
\end{equation}
where $\hat{H}_0$ is the Hamiltonian of the system and $\hat{H}_\text{diss} = - i \gamma \hat{A}^\dagger \hat{A} + \left( \hat{\xi}(t) \hat{A}^\dagger + \hat{A} \hat{\xi}^\dagger(t) \right)$ describes the coupling between the system and the white noise. $\gamma$ is the strength of dissipation and $\hat{A}$ is the dissipative operator. $\hat{\xi}(t)$ is the Langevin noise satisfying $\langle \hat{\xi}(t) \rangle_{\text{noise}} = \langle \hat{\xi}^\dagger(t) \rangle_{\text{noise}} = 0, \langle \hat{\xi}(t) \hat{\xi}(t') \rangle_{\text{noise}} = \langle \hat{\xi}^\dagger(t) \hat{\xi}^\dagger(t') \rangle_{\text{noise}} = 0, \langle \hat{\xi}^\dagger(t) \hat{\xi}(t') \rangle_{\text{noise}} = 0, \langle \hat{\xi}(t) \hat{\xi}^\dagger(t') \rangle_{\text{noise}} = 2 \gamma \delta(t - t')$, which ensure the canonical commutation relation and the conservation of probability. For an observable $\hat{O}$, the relation between the Heisenberg picture and the interaction picture is 
\begin{equation}
    \langle \hat{O}_\text{H}(t) \rangle = \langle \hat{U}^\dagger(t) \hat{O}_\text{I}(t) \hat{U}(t) \rangle,
\end{equation}
where the time evolution operator $\hat{U}(t) = \hat{\mathcal{T}} e^{- i \int_{0}^t dt' \hat{H}_\text{diss, I}(t')}$. For a small $\gamma$, the time evolution operator can be expanded to the first order and substituted into the above equation, then we obtain
\begin{footnotesize}
    \begin{equation}
        \delta \langle \hat{O}(t) \rangle = \gamma \int_0^t dt' \left[ 2 \langle \hat{A}_\text{I}^\dagger(t') \hat{O}_\text{I}(t) \hat{A}_\text{I}(t') \rangle - \langle \{ \hat{A}_\text{I}^\dagger(t') \hat{A}_\text{I}(t'), \hat{O}_\text{I}(t) \} \rangle \right],
    \end{equation}
\end{footnotesize}
which is just the non-hermitian linear response theory~\cite{Pan2020Non}. Combining this non-hermitian linear response theory and the law of error propagation, the estimated precision for $\gamma$ can be written as 
\begin{equation}
    \small \Delta \gamma = \frac{\langle \Delta \hat{O}_\text{H}(t) \rangle}{\left|\int_0^t dt' \left[2 \langle \hat{A}_\text{I}^\dagger(t') \hat{O}_\text{I}(t) \hat{A}_\text{I}(t') \rangle - \langle \{ \hat{A}_\text{I}^\dagger(t') \hat{A}_\text{I}(t'), \hat{O}_\text{I}(t) \} \rangle \right]\right|}. 
\end{equation}
In precision measurement, we usually neglect the evolution of systems under $\hat{H}_0$. There are two aspects to interpret it. One aspect is that $\hat{H}_0$ is indeed zero such as Bell states for photons or some entangled states after we transfer one of two components to an auxiliary state~\cite{Kwiat1995New, Wei2007Hyperentangled, Shwa2013Heralded, DeMille2020Quantum}. Another aspect is that the time evolution operator $e^{- i \int_0^t dt' \hat{H}}$ can be approximately split into two consecutive parts $e^{- i \int_0^t dt \hat{H}_\text{diss}} e^{- i \int_0^t dt \hat{H}_0}$ during a short time, such that the dissipation process can be interpreted as that it is encoded in the state $|i'\rangle = e^{- i \int_0^t dt \hat{H}_0} |i\rangle$ with $|i\rangle$ the initially input state~\cite{Hatano2005Finding, Childs2021Theory}. Eventually, the estimated precision for $\gamma$ is significantly simplified as 
\begin{equation}
    \Delta \gamma = \frac{\langle \Delta \hat{O}_\text{H}(t) \rangle}{\left| 2 \langle \hat{A}^\dagger \hat{O} \hat{A} \rangle - \langle \{ \hat{A}^\dagger \hat{A}, \hat{O} \} \rangle \right| t}.
    \label{eq:Delta_gamma}
\end{equation}

\begin{table}[t]
    \begin{tabular}{|l|l|l|}
    \hline
    \diagbox{Precision}{Fluctuation} & $\langle \Delta \hat{O} \rangle = 0$ & $\langle \Delta \hat{O} \rangle \neq 0$ \\ \hline
    $\qquad \qquad \Delta \gamma$ & $\sqrt{\gamma / 2 \langle \hat{A}^\dagger \hat{A} \rangle t}$ & $1 / r \langle \hat{A}^\dagger \hat{A} \rangle t$ \\ \hline
    \end{tabular}
    \caption{The lower bound of estimated precision for the parameter $\gamma$ with the non-unitary interferometer. If there is no fluctuation of observable $\hat{O}$ for the initial state, $\Delta \gamma$ is bounded by $\sqrt{\gamma / 2 \langle \hat{A}^\dagger \hat{A} \rangle t}$; while if there is fluctuation of observable $\hat{O}$ for the initial state, $\Delta \gamma$ is bounded by $1 / r \langle \hat{A}^\dagger \hat{A} \rangle t$ with $r$ a constant value independent of particle number and evolution time.}
    \label{tab:bound}
\end{table}

According to the above equation, we first analyze the scaling of estimated precision for the parameter $\gamma$ with time $t$. The numerator in Eq.~(\ref{eq:Delta_gamma}) can also be expanded to the linear order of $t$, 
\begin{equation}
  \begin{aligned}
    \langle \Delta \hat{O}_\text{H}(t) \rangle^2 =& \langle \Delta \hat{O} \rangle^2 + \left[2 \langle \hat{A}^\dagger \hat{O}^2 \hat{A} \rangle - \langle \{ \hat{A}^\dagger \hat{A}, \hat{O}^2 \} \rangle \right] \gamma t \\
    & - \left[ 4 \langle \hat{O} \rangle \langle \hat{A}^\dagger \hat{O} \hat{A} \rangle - 2 \langle \hat{O} \rangle \langle \{ \hat{A}^\dagger \hat{A}, \hat{O} \} \rangle \right] \gamma t.
  \end{aligned}
  \label{eq:numerator}
\end{equation}
So, I) if the fluctuation of observable $\hat{O}$ for the initial state is $0$, i.e., $\langle \Delta \hat{O} \rangle = 0$, $\Delta \gamma$ is proportional to $1 / \sqrt{t}$; II) if $\langle \Delta \hat{O} \rangle \neq 0$, $\Delta \gamma$ is proportional to $1 / t$.

Second, we analyze the scaling of estimated precision for the parameter $\gamma$ with particle number $N$. The denominator in Eq.~(\ref{eq:Delta_gamma}) is bounded by 
\begin{equation}
    \begin{aligned} 
        &|2 \langle \hat{A}^\dagger \hat{O} \hat{A} \rangle - \langle \{ \hat{A}^\dagger \hat{A}, \hat{O} \} \rangle| t \\ 
        \leq & |\langle \hat{A}^\dagger [\hat{O}, \hat{A}] \rangle| + |\langle [\hat{A}^\dagger, \hat{O}] \hat{A} \rangle| t \\
        \leq & 2 \sqrt{\langle \hat{A}^\dagger \hat{A} \rangle} \sqrt{\langle [\hat{A}^\dagger, \hat{O}] [\hat{O}, \hat{A}] \rangle} t,
    \end{aligned} 
    \label{eq:denominator}
\end{equation}
where we have used the triangular inequality in the second line and the Schwartz inequality in the third line. So, I) if $\langle \Delta \hat{O} \rangle = 0$, $\langle \Delta \hat{O}_\text{H}(t) \rangle$ can be further simplified as $\langle \Delta \hat{O}_\text{H}(t) \rangle = \sqrt{2 \langle [\hat{A}^\dagger, \hat{O}] [\hat{O}, \hat{A}] \rangle \gamma t}$, then $\Delta \gamma \geq \sqrt{\gamma / 2 \langle \hat{A}^\dagger \hat{A} \rangle t}$, which is proportional to $1 / \sqrt{N}$ or $1 / N$ depending on both of dissipative operators and initial states; II) if $\langle \Delta \hat{O} \rangle \neq 0$, $\hat{O}$ introduces a factor of $N$ because $\hat{O}$ is a collective operator of systems while each commutation relation reduces a factor of $N$~\cite{Watanabe2015Absence}, then $\Delta \gamma \gtrsim \langle \Delta \hat{O} \rangle / r \langle \hat{A}^\dagger \hat{A} \rangle t$ with $r$ a real number independent of particle number and time. Additionally, if $\langle \Delta \hat{O} \rangle \sim 1$, then $\Delta \gamma \gtrsim 1 / r \langle \hat{A}^\dagger \hat{A} \rangle t$, which is proportional to $1 / N$ or $1 / N^2$. 

In general, the estimated precision for the parameter $\gamma$ is bounded by $1 / (\langle \hat{A}^\dagger \hat{A} \rangle t)^\alpha$, which is related to the correlation of the encoding dissipative operator and the evolution time. These results are summarized in Table~\ref{tab:bound}. Furthermore, to demonstrate the correctness and the utility of the above analytical results, we calculate the estimated precision for three different kinds of non-unitarily encoding operators, including particle loss, relaxation, and dephasing.

\section{Examples}

\subsection{Particle loss}

\begin{figure}
    \centering
    \includegraphics[width=8.5cm]{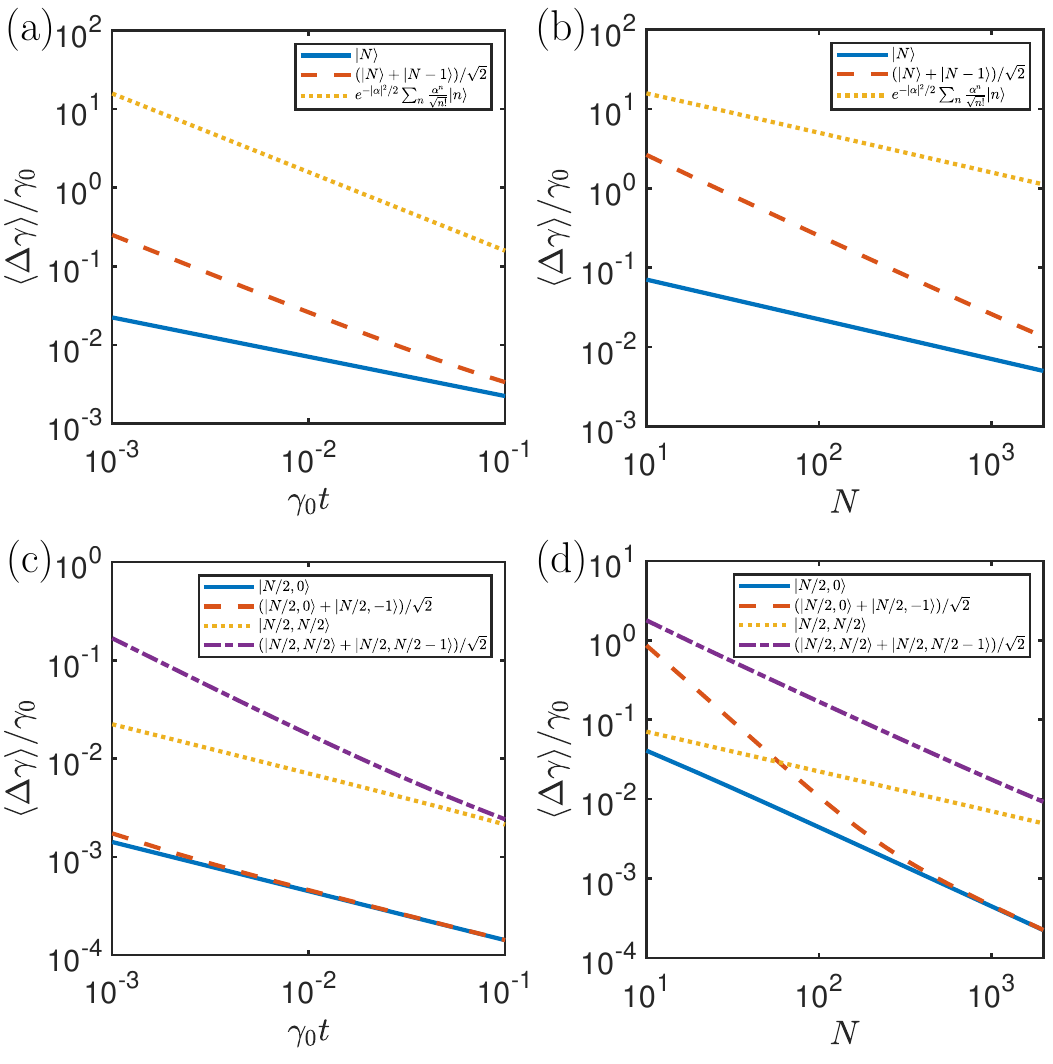}
    \caption{Scaling of estimated precision $\Delta \gamma$ with time $t$ and particle number $N$. (a, b) Estimated precision $\Delta \gamma$ under particle loss for three different initial states, i.e., Fock state $|N\rangle$, critical state $|\psi\rangle = (|N\rangle + |N - 1\rangle) / \sqrt{2}$, and coherent state $|\psi\rangle = e^{- |\alpha|^2 / 2} \sum_n \frac{\alpha^n}{\sqrt{n!}} |n\rangle$. (c, d) Estimated precision $\Delta \gamma$ under relaxation for different initial states, i.e., Dicke state $|N / 2, 0\rangle$, superposition Dicke state $(|N / 2, 0\rangle + |N / 2, - 1\rangle) / \sqrt{2}$, polarized spin state $|N / 2, N / 2\rangle$, and superposition polarized spin state $(|N / 2, N / 2\rangle + |N / 2, N / 2 - 1\rangle) / \sqrt{2}$. $N = 1000$ for (a, c), while $\gamma_0 t = 0.01$ for (b, d). $\gamma / \gamma_0 = 0.001$. The dissipative strength and time units are $\gamma_0$ and $\gamma_0^{- 1}$, respectively.} 
    \label{fig:loss_relaxation}
\end{figure}

We here first consider the one-body loss $\hat{A} = \hat{a}$, and the observable is chosen as the particle number $\hat{O} = \hat{n}$. After substituting these operators into Eq.~\eqref{eq:Delta_gamma}, thus the estimated precision for $\gamma$ becomes
\begin{equation}
    \Delta \gamma = \frac{\sqrt{\langle \Delta \hat{n} \rangle^2 + \left[ 2 \langle \hat{n} \rangle - 4 \langle \Delta \hat{n} \rangle^2 \right] \gamma t}}{|2 \langle \hat{n} \rangle| t}.
\end{equation}
The initial state is chosen as the ground state at different regimes for the Bose-Hubbard model. If the initial state is a Fock state $|\psi\rangle = |N\rangle$, as the ground state of a Mott insulator, $\Delta \gamma = \sqrt{\gamma / 2 N t}$, which reaches the lower bound for the condition $\langle \Delta \hat{n} \rangle = 0$ shown in Table~\ref{tab:bound}. If the initial state is a coherent state $|\psi\rangle = e^{- |\alpha|^2 / 2} \sum_n \frac{\alpha^n}{\sqrt{n!}} |n\rangle$, as the ground state of a superfluid, $\Delta \gamma = 1 / 2 \sqrt{N} t$, which obviously does not reach the lower bound because of $\langle \Delta \hat{n} \rangle = \sqrt{N}$ and $\langle \hat{n} \rangle = N$. However, if the initial state is $|\psi\rangle = (|N\rangle + |N - 1\rangle) / \sqrt{2}$, as the ground state in the critical regime near the superfluid-to-Mott-insulator transition~\cite{Krauth1992Gutzwiller}, $\Delta \gamma = 1 / 2 (2 N - 1) t$, which recovers the lower bound for the condition $\langle \Delta \hat{n} \rangle \neq 0$. The numerical results for $\Delta \gamma$ calculated based on the law of error propagation for these different initial states are shown in Fig.~\ref{fig:loss_relaxation} (a) and (b). In Fig.~\ref{fig:loss_relaxation} (a) and (b), we show that $\Delta \gamma$ decreases as time $t$ and particle number $N$ increase, and the scaling of numerical results is exactly consistent with our above analysis.

\subsection{Relaxation}

Second, we consider the relaxation $\hat{A} = \hat{J}_-$, and the observable is chosen as the angular momentum $\hat{O} = \hat{J}_z$. We also substitute these operators into Eq.~\eqref{eq:Delta_gamma}, then the estimated precision for $\gamma$ is
\begin{equation}
    \small \Delta \gamma = \frac{\sqrt{\langle \Delta \hat{J}_z \rangle^2 + \left[ 2 \langle \hat{J}_+ \hat{J}_- \rangle - 4 \langle \hat{J}_+ \hat{J}_- \hat{J}_z \rangle + 4 \langle \hat{J}_+ \hat{J}_- \rangle \langle \hat{J}_z \rangle  \right] \gamma t}}{|2 \langle \hat{J}_+ \hat{J}_- \rangle| t}.
\end{equation}
If the initial state is a polarized spin state $|\psi\rangle = |j = N / 2, m = N / 2\rangle$ without entanglement, $\Delta \gamma = \sqrt{\gamma / 2 N t}$, while if the initial state is a Dicke state $|\psi\rangle = |N / 2, 0\rangle$ with high entanglement~\cite{Dicke1954Coherence}, $\Delta \gamma = \sqrt{\gamma / N(N/2 + 1)t}$. Both of the above two states reach the lower bound for the condition $\langle \Delta \hat{J}_z \rangle = 0$. Here we note that the estimated precision for the parameter $\gamma$ is proportional to $1 / N$ if the initial state is a Dicke state, which is much smaller than $1 / \sqrt{N}$ and significantly enhances the metrology when $N$ is large. This stands to reason that the response is extremely fast resulting from superradiance and the entanglement indeed enhance the metrology even for estimating the dissipative parameter. We can also rotate the above two states along the $y$-axis as the initial states $e^{- i \hat{J}_y \theta} |\psi\rangle$ that must be superposition states with different magnetizations. To ensure the initial fluctuation $\langle \Delta \hat{J}_z \rangle \sim 1$, without loss of generality, we choose one superposition state as $(|N / 2, N / 2\rangle + |N / 2, N / 2 - 1\rangle) / \sqrt{2}$, then $\Delta \gamma = 1 / 2 (3 N - 2) t$, while we choose another superposition state as $(|N / 2, 0\rangle + |N / 2, - 1\rangle) / \sqrt{2}$, then $\Delta \gamma = 1 / (N^2 + 2 N - 4) t$. They also reach the lower bound for the condition $\langle \Delta \hat{J}_z \rangle \neq 0$. 

The numerical results for $\Delta \gamma$ corresponding to these four different initial states under relaxation based on the law of error propagation are shown in Fig.~\ref{fig:loss_relaxation} (c) and (d). According to Fig.~\ref{fig:loss_relaxation} (c), we find the scaling of $\Delta \gamma$ with time $t$ is $- 1$ for the initial state $(|N / 2, 0\rangle + |N / 2, - 1\rangle) / \sqrt{2}$ during a very short time, then it quickly approaches $- 1 / 2$ that is the same as the scaling for the Dicke state $|N / 2, 0\rangle$. Besides, there is a similar result that the scaling for the initial state $(|N / 2, N / 2\rangle + |N / 2, N / 2 - 1\rangle) / \sqrt{2}$ finally approaches that for the polarized spin state $|N / 2, N / 2\rangle$. These phenomena are owing to the fluctuation of the observable increasing as time goes on such that the initial fluctuation does not dominate. The scaling for the superposition Dicke state approaches that of the Dicke state faster than the superposition polarized spin state approaches the polarized spin state, since the fluctuation of observable for a superradiant Dicke state increases faster. In Fig.~\ref{fig:loss_relaxation} (d), the scaling of $\Delta \gamma$ with total particle number $N$ is also consistent with the above analysis.

\subsection{Dephasing}

\begin{figure}
    \centering
    \includegraphics[width=8.5cm]{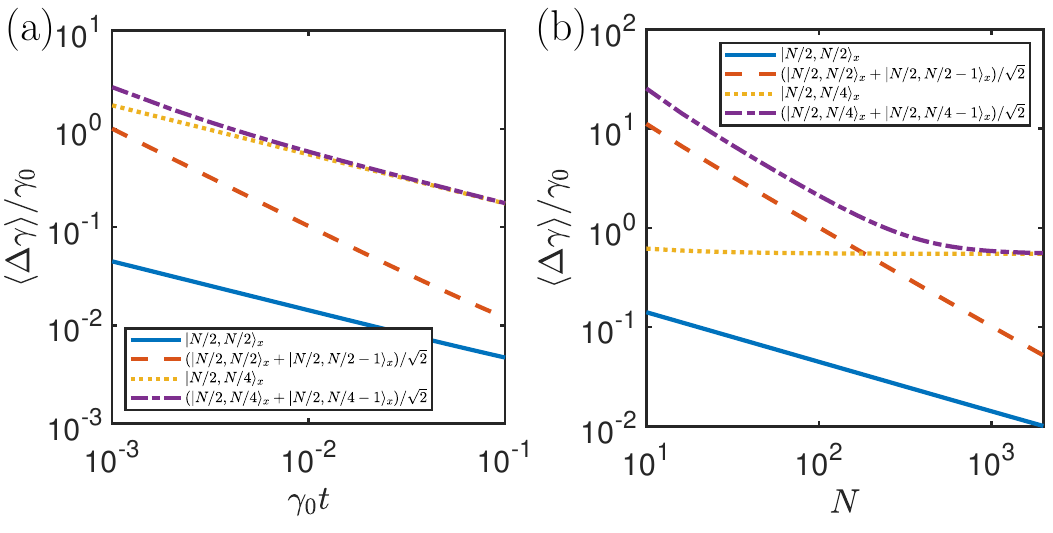}
    \caption{Scaling of estimated precision $\Delta \gamma$ with time $t$ (a) and particle number $N$ (b) under dephasing. Different lines correspond to different initial states labeled in the legend. $N = 1000$ for (a), while $\gamma_0 t = 0.01$ for (b). $\gamma / \gamma_0 = 0.001$. The dissipative strength and time units are $\gamma_0$ and $\gamma_0^{- 1}$, respectively.} 
    \label{fig:dephasing}
\end{figure}

Third, we consider the dephasing $\hat{A} = \hat{J}_z$, and the observable is chosen as $\hat{O} = \hat{J}_x$. According to Eq.~\eqref{eq:Delta_gamma}, the estimated precision for $\gamma$ is written as 
\begin{equation}
    \Delta \gamma = \frac{\sqrt{\langle \Delta \hat{J}_x \rangle^2 + \left[ 2 \langle \hat{J}_y^2 \rangle - 2 \langle \Delta \hat{J}_x \rangle^2 \right] \gamma t}}{|\langle \hat{J}_x \rangle| t}.
\end{equation}
If the initial state is a polarized spin state $|\psi\rangle = |N / 2, N / 2\rangle_x$, where the subscript $x$ represents the quantization axis, then $\Delta \gamma = \sqrt{2 \gamma / N t}$, which reaches the lower bound for the condition $\langle \Delta \hat{J}_x \rangle = 0$. If the initial state is a superposition state $|\psi\rangle = (|N / 2, N / 2\rangle_x + |N / 2, N / 2 - 1\rangle_x) / \sqrt{2}$, then $\Delta \gamma = 1 / (N - 1) t$, which also reaches the lower bound for the condition $\langle \Delta \hat{J}_x \rangle \neq 0$. If the initial states are highly entangled states, such as $|N / 2, N / 4\rangle_x$ or $(|N / 2, N / 4\rangle_x + |N / 2, N / 4 - 1\rangle_x) / \sqrt{2}$, we obtain $\Delta \gamma = \sqrt{3 \gamma / t}$ or $\Delta \gamma = 2 / (N - 2) t$, respectively. However, according to Table.~\ref{tab:bound}, the lower bound for these two highly entangled states are calculated as $\sqrt{\gamma / N^2 t}$ or $1 / r N^2 t$, respectively. Obviously, the estimated precision for these highly entangled states do not reach the lower bound. This is not due to inappropriately initial states like a coherent state under particle loss, but due to overestimating the denominator in Eq.~\eqref{eq:denominator} for this dissipative channel. Here we must stress that the dephasing operator is a hermitian operator while the operators of relaxation and particle loss are non-hermitian. For a hermitian dissipative operator, the denominator becomes
\begin{equation}
    |2 \langle \hat{A} \hat{O} \hat{A} \rangle - \langle \{ \hat{A} \hat{A}, \hat{O} \} \rangle| t = |\langle [\hat{A}, [\hat{O}, \hat{A}]] \rangle| t.
\end{equation}
There exists an extra commutation relation compared with Eq.~(\ref{eq:denominator}), and each commutation relation reduces a factor of $N$. So, the inequality in Eq.~(\ref{eq:denominator}) should be modified and is constrained much tighter. In general, when there is an entangled state such that the correlation $\langle \hat{A}^\dagger \hat{A} \rangle \sim N^2$, the lower bound shown in Table.~\ref{tab:bound} will overestimate the scaling with particle number $N$. It should be modified as $N \sqrt{\gamma / 2 \langle \hat{A}^\dagger \hat{A} \rangle t}$ for the condition $\langle \Delta \hat{O} \rangle = 0$, while $N / r \langle \hat{A}^\dagger \hat{A} \rangle t$ for the condition $\langle \Delta \hat{O} \rangle \neq 0$. The numerical results for $\Delta \gamma$ based on the law of error propagation for different initial states under dephasing are shown in Fig.~\ref{fig:dephasing}. These numerical results further demonstrate that the scaling for highly entangled states is modified correctly.

\section{Discussions}

\subsection{Estimated precision $\Delta \gamma$ for a long time}

\begin{figure}[t]
    \centering
    \includegraphics[width=8.8cm]{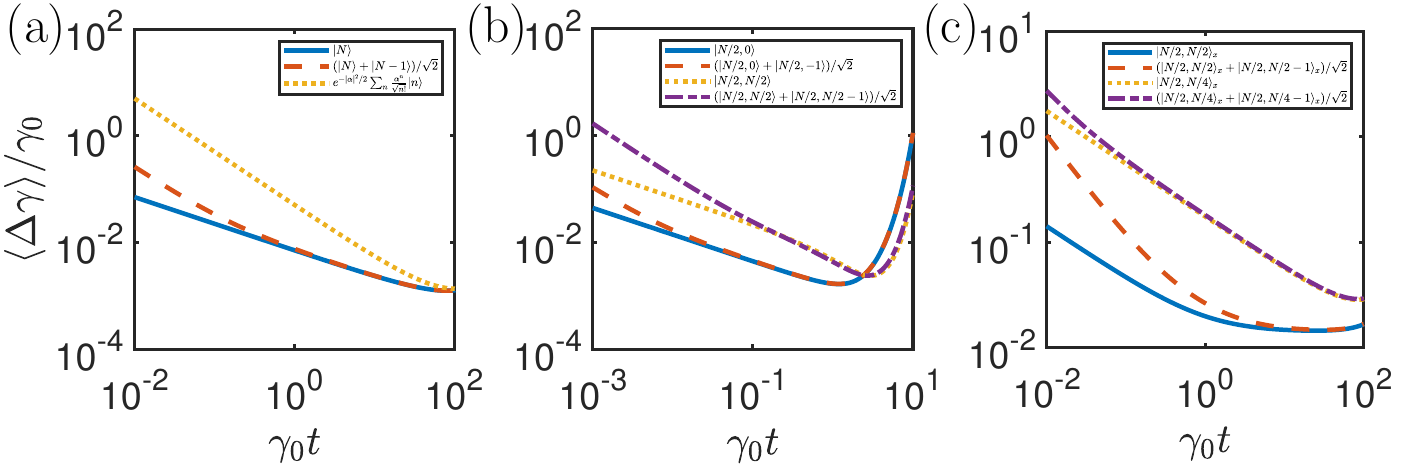}
    \caption{The estimated precision $\Delta \gamma$ with time $t$ for different initial states under particle loss (a), relaxation (b), and dephasing (c). Different lines correspond to different initial states labeled in the legend. The total particle number $N = 100$, and the strength of dissipation $\gamma / \gamma_0 = 0.01.$} 
    \label{fig:long_time}
  \end{figure}

According to the analytical solutions in Table.~\ref{tab:bound}, a natural question is whether the estimated precision $\Delta \gamma$ for $\langle \Delta \hat{O} \rangle \neq 0$ will be smaller than that for $\langle \Delta \hat{O} \rangle = 0$ after a long time because the scaling of $\Delta \gamma$ with time $t$ is $- 1$ for $\langle \Delta \hat{O} \rangle \neq 0$ and $- 1 / 2$ for $\langle \Delta \hat{O} \rangle = 0$. To clarify this question, we numerically calculate $\Delta \gamma$ for a long time, and the results are shown in Fig.~\ref{fig:long_time}. According to these results, we find $\Delta \gamma$ for $\langle \Delta \hat{O} \rangle \neq 0$ will approach that for $\langle \Delta \hat{O} \rangle = 0$ but is not smaller than that for $\langle \Delta \hat{O} \rangle = 0$. This can be understood easily based on the law of error propagation. Here we discuss a dephasing process as an example. The denominator $\partial \langle \hat{O}(t) \rangle / \partial \gamma$ is almost the same for initial states $|N / 2, N / 2\rangle$ and $(|N / 2, N / 2\rangle + |N / 2, N / 2 - 1\rangle) / \sqrt{2}$ while the numerator $\langle \Delta \hat{O} \rangle = 0$ for $|N / 2, N / 2\rangle$ and $\langle \Delta \hat{O} \rangle = 1 / 2$ for $(|N / 2, N / 2\rangle + |N / 2, N / 2 - 1\rangle) / \sqrt{2}$. If $\Delta \gamma$ is as small as possible, the numerator should also be as small as possible. So, the estimated precision $\Delta \gamma$ for $|N / 2, N / 2\rangle$ is smaller than that for $(|N / 2, N / 2\rangle + |N / 2, N / 2 - 1\rangle) / \sqrt{2}$ during a short time, then when the fluctuation of the numerator increases as time goes on such that the initial fluctuation does not dominate and the fluctuations for these two states are almost the same, the estimated precision for these two initial states will merge. In general, the fact that $\Delta \gamma$ for $\langle \Delta \hat{O} \rangle = 0$ is smaller than that for $\langle \Delta \hat{O} \rangle \neq 0$ is irrespective of the scaling. Furthermore, we find $\Delta \gamma$ increases after a enough long time, which is attributed to almost no particle in the particle loss process, almost no spin flip in the relaxation process, and almost no coherence in the dephasing process. So, the effective information extracted from the measurement will decrease after a enough long time. Besides, the above analysis about the scaling is suitable for $\Delta \gamma$ for a large particle number $N$.

\subsection{Relation between the lower bound and the quantum Fisher information}

\begin{figure}[t]
    \centering
    \includegraphics[width=8.8cm]{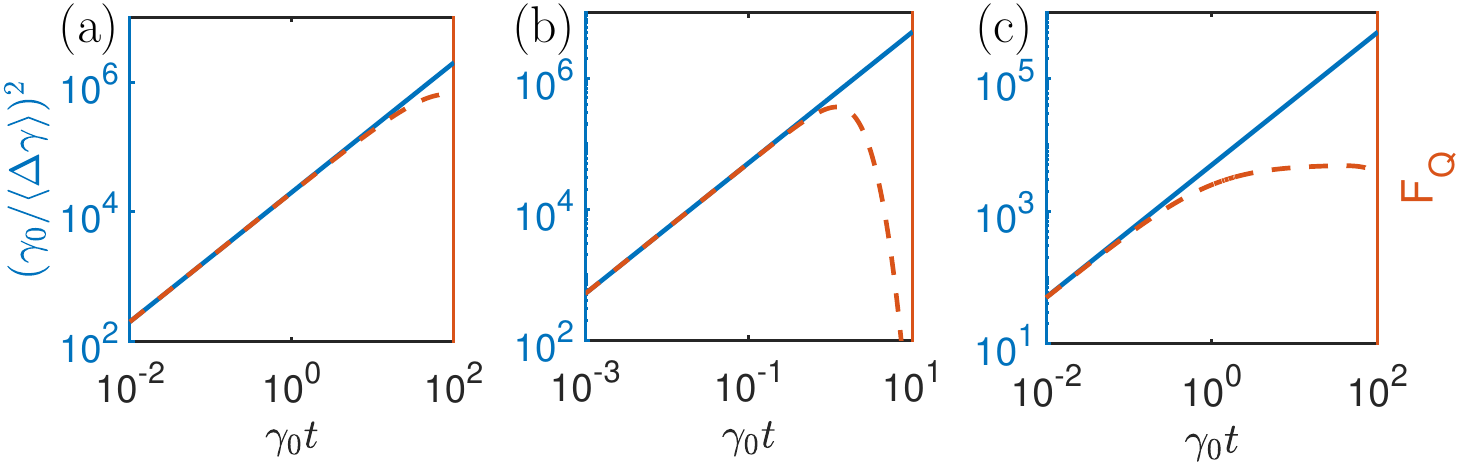}
    \caption{The estimated precision $\Delta \gamma$ with time $t$ for initial Fock state under particle loss (a), initial Dicke state under relaxation (b), and initial polarized state under dephasing (c). The left $y$-axis represents $(\gamma_0 / \Delta \gamma)^2$, and the right $y$-axis represents the quantum Fisher information. The tick labels of double $y$-axes are the same. The total particle number $N = 100$, and the strength of dissipation $\gamma / \gamma_0 = 0.01.$} 
    \label{fig:fisher_information}
\end{figure}

For a closed system, we know the inverse of the square of the lower bound, i.e., $4 \langle \Delta \hat{A} \rangle^2 t^2$, is equal to the quantum Fisher information, as demonstrated in Appendix.~\ref{sec:app1}. For an open system, we also compare the inverse of the square of the lower bound, i.e., $2 \left\langle \hat{A}^\dagger \hat{A} \right\rangle t / \gamma$, with the quantum Fisher information~\cite{Braunstein1994Statistical}, and the numerical results are shown in Fig.~\ref{fig:fisher_information}. Combining the numerical results in Fig.~\ref{fig:long_time}, we find that $2 \left\langle \hat{A}^\dagger \hat{A} \right\rangle t / \gamma$ is indeed consistent with the quantum Fisher information when the non-hermitian linear response theory is satisfied. Otherwise, the prediction of $2 \left\langle \hat{A}^\dagger \hat{A} \right\rangle t / \gamma$ will overestimate the information that we can extract from the measurement. Based on the equation $2 \left\langle \hat{A}^\dagger \hat{A} \right\rangle t / \gamma$, we will obtain more and more information as time goes on and even in the limit of infinite time. However, in this limit, we assure that there are no quantum resources extracted from a steady state for an open system, which is consistent with the behavior of the red dashed line in Fig.~\ref{fig:fisher_information}.

\section{Conclusion}

The lower bound of estimated precision for a parameter unitarily encoded in closed systems has been obtained, which is related to the fluctuation of the encoding operator and the evolution time. Here, we demonstrate the estimated precision for the dissipative strength non-unitarily encoded in open systems is bounded by the correlation of the encoding operator based on the non-hermitian linear response theory. Additionally, we explicitly calculate the estimated precision for dissipative parameters corresponding to three different kinds of non-unitarily encoding processes, including particle loss, relaxation, and dephasing. These results further confirm the lower bound. Finally, we discuss the relation between the lower bound and the quantum Fisher information, and they are consistent before failure of non-hermitian linear response theory. May there are other interesting questions, for example, what is the lower bound when both unitary and non-unitary encoding exist? What is the lower bound for estimated precision based on the nonlinear response if there is no linear response?

{\bf Acknowledgements} This project is supported by the National Natural Science Foundation of China (Grant No. 12375023 and No. 12204428); the National Key R \& D Program of China (Grants No. 2022YFA1404500 and No. 2021YFA1400900); Cross-disciplinary innovative research group project of Henan province (Grant No. 232300421004); the National Natural Science Foundation of China (Grants No. 12125406); the Natural Science Foundation of Henan Province (Grant No.242300421159).

\appendix

\section{Parameter estimation based on the linear response theory}
\label{sec:app1}

Suppose a system is coupled to an external field as follows,
\begin{equation}
    \hat{H} = \hat{H}_0 + \Omega \hat{A},
\end{equation}
where $\hat{H}_0$ is the Hamiltonian of the system before encoding and $\Omega \hat{A}$ is the encoding part. For an observable $\hat{O}$, the relation between the Heisenberg picture and the interaction picture is
\begin{equation}
    \langle \hat{O}_\text{H}(t) \rangle = \langle \hat{U}^\dagger(t) \hat{O}_\text{I}(t) \hat{U}(t) \rangle,
    \label{eq:app1_2}
\end{equation}
where the time evolution operator $\hat{U}(t) = \hat{\mathcal{T}} e^{- i \Omega \int_{0}^t dt' \hat{A}_\text{I}(t')}$. The subscript $\text{H}$ and $\text{I}$ represent the Heisenberg picture and the interaction picture, respectively. For a small $\Omega$, we can expand the time evolution operator to the first order,
\begin{equation}
    \hat{U}(t) \approx 1 - i \Omega \int_{0}^t dt' \hat{A}_\text{I}(t').
    \label{eq:app1_3}
\end{equation}
After we substitute Eq.~(\ref{eq:app1_3}) into Eq.~(\ref{eq:app1_2}), we obtain 
\begin{equation}
    \delta \langle \hat{O}(t) \rangle = \langle \hat{O}_\text{H}(t) \rangle - \langle \hat{O}_\text{I}(t) \rangle 
    = - i \Omega \int_{0}^t dt' \langle [\hat{O}_\text{I}(t), \hat{A}_\text{I}(t')] \rangle,
\end{equation}
then the estimated precision for $\Omega$ can be calculated as
\begin{equation}
    \Delta \Omega = \frac{\langle \Delta \hat{O}_\text{H}(t) \rangle}{|\partial \langle \hat{O}(t) \rangle / \partial \Omega|} = \frac{\langle \Delta \hat{O}_\text{H}(t) \rangle}{|\int_{0}^t dt' \langle [\hat{O}_\text{I}(t), \hat{A}_\text{I}(t')] \rangle|}.
\end{equation}
In precision measurement, we usually neglect the evolution of systems under $\hat{H}_0$, then the above equation becomes 
\begin{equation}
    \Delta \Omega = \frac{\langle \Delta \hat{O}_\text{H}(t) \rangle}{|\langle [\hat{O}, \hat{A}] \rangle| t} \geq \frac{\langle \Delta \hat{O}_\text{H}(t) \rangle}{2 \langle \Delta \hat{O} \rangle \langle \Delta \hat{A} \rangle t} \geq \frac{1}{2 \langle \Delta \hat{A} \rangle t},
\end{equation}
where we have applied the Heisenberg uncertainty relation $\langle \Delta \hat{O} \rangle \langle \Delta \hat{A} \rangle \geq \frac{1}{2} |\langle [\hat{O}, \hat{A}] \rangle|$ in the first inequality, and we have assumed the fluctuation of the observable $\hat{O}$ is the smallest for the initial state in the second inequality. So, the estimated precision for $\Omega$ is bounded by the fluctuation of the encoding operator $\hat{A}$ and time $t$. 

Besides, the quantum Fisher information for closed systems under unitary encoding can be obtained by~\cite{Braunstein1994Statistical}
\begin{equation}
  \text{F}_\Omega = 4 \left[ \left\langle \partial_\Omega \psi(\Omega) | \partial_\Omega \psi(\Omega)  \right\rangle - \left| \left\langle \partial_\Omega \psi(\Omega) | \psi(\Omega) \right\rangle \right|^2 \right],
\end{equation}
where $|\psi(\Omega)\rangle = e^{- i \Omega t \hat{A}} |\psi(0)\rangle$. Thus $\text{F}_\Omega = \left( \Delta \hat{A} t \right)^2$, and the lower bound of estimated precision can be rewritten as 
\begin{equation}
  \Delta \Omega \geq \frac{1}{\sqrt{\text{F}_\Omega}},
\end{equation}
which is consistent with the results obtained in the previous paper~\cite{Braunstein1994Statistical}.

\section{Deriving the Lindblad equation from the non-hermitian Hamiltonian $\hat{H} = \hat{H}_0 + \hat{H}_\text{diss}$}

\begin{widetext}
An open system can be described by a non-hermitian Hamiltonian, where $\hat{H}_0$ is the Hamiltonian of the system and $\hat{H}_\text{diss} = - i \gamma \hat{A}^\dagger \hat{A} + \left( \hat{\xi}(t) \hat{A}^\dagger + \hat{A} \hat{\xi}^\dagger(t) \right)$ describes the coupling between the system and the bath. $\gamma$ is the strength of dissipation and $\hat{A}$ is the dissipative operator. $\hat{\xi}(t)$ is the Langevin noise satisfying
\begin{equation}
    \begin{aligned}
        \langle \hat{\xi}(t) \rangle_{\text{noise}} &= \langle \hat{\xi}^\dagger(t) \rangle_{\text{noise}} = 0, 
        \langle \hat{\xi}(t) \hat{\xi}(t') \rangle_{\text{noise}} = \langle \hat{\xi}^\dagger(t) \hat{\xi}^\dagger(t') \rangle_{\text{noise}} = 0, 
        \langle \hat{\xi}^\dagger(t) \hat{\xi}(t') \rangle_{\text{noise}} = 0, 
        \langle \hat{\xi}(t) \hat{\xi}^\dagger(t') \rangle_{\text{noise}} = 2 \gamma \delta(t - t').
    \end{aligned}
\end{equation}
Considering an observable $\hat{O}(t)$ of the system, its evolution is
\begin{equation}
    \begin{aligned}
        &\hat{O}(t + \delta t) = \hat{U}^\dagger(t, t + \delta t) \hat{O}(t) \hat{U}(t, t + \delta t) \\
        &= \hat{\mathcal{T}} e^{i \int_t^{t + \delta t} \hat{H}^\dagger(t') dt'} \hat{O}(t) \hat{\mathcal{T}} e^{- i \int_t^{t + \delta t} \hat{H}(t^{''}) dt^{''}} \\
        &\approx \left(1 + i \int_t^{t + \delta t} \hat{H}^\dagger(t') dt'\right) \hat{O}(t) \left(1 - i \int_t^{t + \delta t} \hat{H}(t^{''}) dt^{''}\right) \\
        &= \hat{O}(t) + i \int_t^{t + \delta t} [\hat{H}_0, \hat{O}(t)]  dt' - \gamma \int_t^{t + \delta t} \{ \hat{A}^\dagger \hat{A}, \hat{O}(t) \}  dt' + \int_t^{t + \delta t} \int_t^{t + \delta t} \left( \hat{\xi}(t') \hat{A}^\dagger + \hat{A} \hat{\xi}^\dagger(t') \right) \hat{O}(t) \left( \hat{\xi}(t^{''}) \hat{A}^\dagger + \hat{A} \hat{\xi}^\dagger(t^{''}) \right) dt' dt^{''} \\
        &= \hat{O}(t) + i \int_t^{t + \delta t} [\hat{H}_0, \hat{O}(t)] dt' - \gamma \int_t^{t + \delta t} \{ \hat{A}^\dagger \hat{A}, \hat{O}(t) \} dt' + 2 \gamma \int_t^{t + \delta t} \hat{A}^\dagger \hat{O}(t) \hat{A} dt'.
    \end{aligned}
\end{equation}
In the last equality, we have averaged the Langevin noise. Then, we obtain 
\begin{equation}
    \frac{\partial \hat{O}(t)}{\partial t} = i [\hat{H}_0, \hat{O}(t)] - \gamma \{ \hat{A}^\dagger \hat{A}, \hat{O}(t) \} + 2 \gamma \hat{A}^\dagger \hat{O}(t) \hat{A}.
\end{equation}
Furthermore, due to $\partial \text{Tr}(\hat{O}(t) \rho)/\partial t = \partial \text{Tr}(\hat{O} \rho(t))\partial t$ for any observable, we finally obtain the Lindblad equation
\begin{equation}
    \frac{\partial \rho(t)}{\partial t} = - i [\hat{H}_0, \rho(t)] - \gamma \{ \hat{A}^\dagger \hat{A}, \rho(t) \} + 2 \gamma \hat{A} \rho(t) \hat{A}^\dagger.
\end{equation}
\end{widetext}

\normalem
%

\end{document}